\documentclass[conference]{IEEEtran}
\IEEEoverridecommandlockouts

\usepackage{cite}
\usepackage{amsmath,amssymb,amsfonts}
\usepackage{algorithmic}
\usepackage{graphicx}
\usepackage{textcomp}
\usepackage[usenames,dvipsnames]{xcolor}
\def\BibTeX{{\rm B\kern-.05em{\sc i\kern-.025em b}\kern-.08em
    T\kern-.1667em\lower.7ex\hbox{E}\kern-.125emX}}

\usepackage[utf8]{inputenc}
\usepackage[english]{babel}
\usepackage{pgfplots}
\usepackage{amsmath}
\usepackage{amsthm}
\usepackage{amssymb}
\usepackage{bbm}
\usepackage{bm}
\usepackage{braket}
\usepackage{graphicx}
\usepackage{dsfont}
\usepackage[T1]{fontenc}
\usepackage[normalem]{ulem}

\usepackage{subcaption}
\usepackage{ragged2e}

\usetikzlibrary{arrows.meta}

\usepackage[colorlinks=true,citecolor=MidnightBlue,linkcolor=MidnightBlue,urlcolor=MidnightBlue]{hyperref}

\makeatletter
\newcommand*{\centerfloat}{%
	\parindent \z@
	\leftskip \z@ \@plus 1fil \@minus \textwidth
	\rightskip\leftskip
	\parfillskip \z@skip}
\makeatother

\usepackage[capitalize]{cleveref}

\begin{document}

\title{Distributing graph states with a photon-weaving quantum server
\thanks{D.B. was supported by the JST Moonshot R\&D program under Grant JPMJMS226C.}
}

\author{\IEEEauthorblockN{Daniel Bhatti}
\IEEEauthorblockA{\textit{Networked Quantum Devices Unit} \\
\textit{Okinawa Institute of Science and Technology Graduate University}\\
Okinawa, Japan \\
daniel.bhatti@fau.de}
\and
\IEEEauthorblockN{Kenneth Goodenough}
\IEEEauthorblockA{\textit{College of Information and Computer Sciences} \\
\textit{University of Massachusetts Amherst}\\
Amherst, Massachusetts 01003, USA \\
kdgoodenough@gmail.com}
}

\maketitle

\begin{abstract}
One of the key aims of quantum networks is the efficient distribution of multipartite entangled states among end users. While various architectures have been proposed, each comes with distinct advantages and limitations. Many designs depend on long-lived quantum memories and deterministic gates, which, while powerful, introduce considerable cost and technical challenges. Experimentally cheaper alternatives that circumvent these constraints are often limited to specific types of entanglement and a specific number of users. Here, we present an experiment-friendly quantum server that relies only on linear optical elements, offering a flexible approach to multipartite entanglement distribution. Our so-called \textit{photon-weaving quantum server} can generate and distribute one of several locally nonequivalent graph states, including Greenberger-Horne-Zeilinger (GHZ) states, as well as path, cycle, and caterpillar graph states. This is achieved through two distinct fusion protocols, i.e., multiphoton graph-state fusion (\textit{graph-state weaving}) and multiphoton GHZ-state fusion (\textit{GHZ-state weaving}), and can readily be implemented.
\end{abstract}

\begin{IEEEkeywords}
Quantum Network,
Quantum Server,
Graph States,
Quantum Optics,
Photonic Fusion
\end{IEEEkeywords}

\section{Introduction}

Quantum applications often require the generation and distribution of multipartite entangled quantum states in a quantum network~\cite{Murta2020,Raussendorf2003,Kang2024,Meyer2024,Bartolucci2023,Wallnofer2016, li2024generalized,Frantzeskakis2023}.
While one is typically limited to a fixed network topology,
many applications benefit from---or even require---different types of entanglement.
For example, Greenberger-Horne-Zeilinger (GHZ) states are ideal for quantum communication~\cite{Murta2020}, and cluster states are ideal for measurement-based quantum computing (MBQC)~\cite{Raussendorf2003}.

Interestingly, both GHZ states (up to single-qubit rotations) and cluster states belong to one of the most essential types of entangled states, namely graph states. Graph states are quantum states that can be described by a mathematical graph~\cite{hein2004multiparty}, and have a wide range of quantum applications~\cite{Murta2020,Raussendorf2003,Kang2024,Meyer2024,Bartolucci2023,Wallnofer2016,li2024generalized,Frantzeskakis2023}. Besides quantum communication and MBQC, other useful examples are long-range quantum teleportation~\cite{Kang2024}, self-testing~\cite{Meyer2024}, fusion and repeater protocols~\cite{Bartolucci2023,Wallnofer2016, li2024generalized}, and extracting higher-entangled states from noisy states~\cite{Frantzeskakis2023}. It is, therefore, important to design flexible quantum networks that are experimentally feasible but can still distribute various entangled states~\cite{Hahn2019,Meignant2019}.

In general, there are different approaches to constructing quantum networks capable of distributing (multipartite) entanglement~\cite{Li2024}.
While each approach has its own drawbacks and benefits, one promising building block for more complex network topologies is a star-shaped network with a central node, i.e., a \emph{quantum server} that shares an entangled Bell pair with each of the user nodes~\cite{avis2023analysis}.
Now, one could assume long-lived memories, which are essential for storing, manipulating, and entangling qubits, and therewith the capability to perform deterministic single- and two-qubit operations. This would allow for designing powerful quantum servers, i.e., quantum servers that can distribute various types of entangled states~\cite{Caprara2019,Nain2020,Bugalho2023,avis2023analysis,Mazza2025, sen2023multipartite}.
However, such devices might not always be practical since they require sophisticated experimental setups~\cite{avis2023analysis}.

Several proposals have investigated ways around the above-mentioned need for quantum memories.
Following the star-shaped network topology, one simple alternative is a quantum server directly distributing multiphoton entangled states to the distinct users. Possible distribution schemes could utilize postselection~\cite{Kumar2023,Meyer-Scott2022,Bhatti2023}, heralding~\cite{Cao2024,Chen2024,Bhatti2025}, or deterministic photon emitters~\cite{Hilaire2023,Thomas2024}. A second alternative is a quantum server based on linear optical elements and performing multiphoton entanglement swapping~\cite{Wang2009,Grasselli2019,Caprara2019,Roga2023,Ainley2024,Shimizu2025,Zo2025}. So far, such schemes have been discussed for the generation of W states~\cite{Grasselli2019,Ainley2024}, Dicke states~\cite{Roga2023}, and GHZ states~\cite{Wang2009,Caprara2019,Shimizu2025,Ainley2024,Zo2025}. However, a flexible, i.e., experiment-friendly, quantum server that can flexibly decide to distribute one out of several graph states based on linear optics is still missing.

Here, we propose such a quantum server based on linear optical elements only, which generates different locally nonequivalent graph states by i) choosing one out of two photonic fusion protocols, i.e., multiphoton graph-state fusion called \textit{graph-state weaving}~\cite{Lin2011} and multiphoton GHZ-state fusion~\cite{Sagi2003}, which we call \textit{GHZ-state weaving}; ii) performing different single qubit measurements in different bases to manipulate the distributed graph states further.
This so-called \textit{photon-weaving quantum server} (PWQS) is capable of distributing GHZ states, as well as path, cycle, and caterpillar graph states (see ~\cite{Thomas2024, chin2024boson, lobl2024generating, coiteux2024genuinely, de2020protocols} for a nonexhaustive list of references on their generation, distribution, and applications). Notably, these distinct classes of states can be produced using the same experimental setup, requiring only modifications to single-photon operations. Consequently, our PWQS offers an experimentally efficient platform for the distribution of multiple key classes of entangled states.

The paper is structured as follows. First, we describe the theoretical background on graph states and photon-weaving in Section~\ref{sec:background}. In Section~\ref{sec:no_storage}, we discuss our PWQS without the possibility of storing photons, i.e., all photons have to be fused and measured simultaneously. Afterward, in Section~\ref{sec:storage}, we discuss an adjusted version of our PWQS with the possibility of storing photons, which generates small building blocks connecting one or two users first. 
The building blocks are then fused in a second step. We classify the resultant graph states that a PWQS can generate in Appendix~\ref{sec:zigzag_vertex_minors}.  We conclude with an outlook for future work in \cref{sec:Conclusion}.

\section{Background}
\label{sec:background}

In this section, we review the basic concepts and tools we make use of throughout this work.

\subsection{Graph States}

A graph state $\ket{\mathcal{G}}$ is a quantum state that is represented by a graph $\mathcal{G}=(V,E)$, given by a set of $N$ vertices $V$ and a set of $M$ edges $E$~\cite{hein2004multiparty}.
In linear optical implementations, the $N$ vertices correspond to the $N$ photons $v_{1},v_{2},\ldots,v_{N}$ and the $M$ edges $(v_{i},v_{j})\in E$ indicate the $M$ entangling operations performed between the qubits to generate the graph state. That is, to generate an $N$-photon graph state, one applies the controlled Pauli Z (CZ) gates according to the edges in $\mathcal{G}$ to $N$ photons, each prepared in the state $\ket{+}$. Graph states can, therefore, be written in the form
\begin{align}
    \ket{\mathcal{G}} = \prod_{(v_i,v_j)\in E} \text{CZ}_{(v_i,v_j)}\ket{+}\ket{+}\cdots\ket{+}.\label{eq:graph_state}
\end{align}

Graph states are a subset of stabilizer states. In fact, every stabilizer state is equivalent to a graph state, up to single-qubit Cliffords~\cite{van2004graphical}. Our focus on graph states---as opposed to stabilizer states---is thus one out of convenience. As an example, the well-known $N$-qubit GHZ state 

\begin{align}
\label{eq:GHZ}
    \ket{\text{GHZ}_{N}} = \frac{1}{\sqrt{2}} \left( \ket{00\ldots0} + \ket{11\ldots1} \right),    
\end{align}
is single-qubit Clifford equivalent to both a star- and a complete graph state~\cite{hein2004multiparty}. One powerful fact is that two graph states are single-qubit Clifford equivalent if and only if the two underlying graphs are related by a sequence of \emph{local complementations}~\cite{van2004graphical}. A local complementation on a vertex $v$ changes the presence/absence of an edge between each possible pair of neighbors $w_1, w_2$ of $v$. That is, if there is an edge between neighbors $w_1, w_2$ of $v$, remove the edge; if there is no edge between neighbors $w_1, w_2$ of $v$, add the edge. We say that two graphs (and their associated graph states) are locally equivalent if they are related by a sequence of local complementations.

\subsection{Postselection}

Here, we introduce the different linear-optical postselection methods used in this work (further details can be found, e.g., in~\cite{Sagi2003,Pilnyak2017,Lin2011}).

\subsubsection{Postselecting GHZ States}

\begin{figure}
    \centering
    \includegraphics[width=0.9\linewidth]{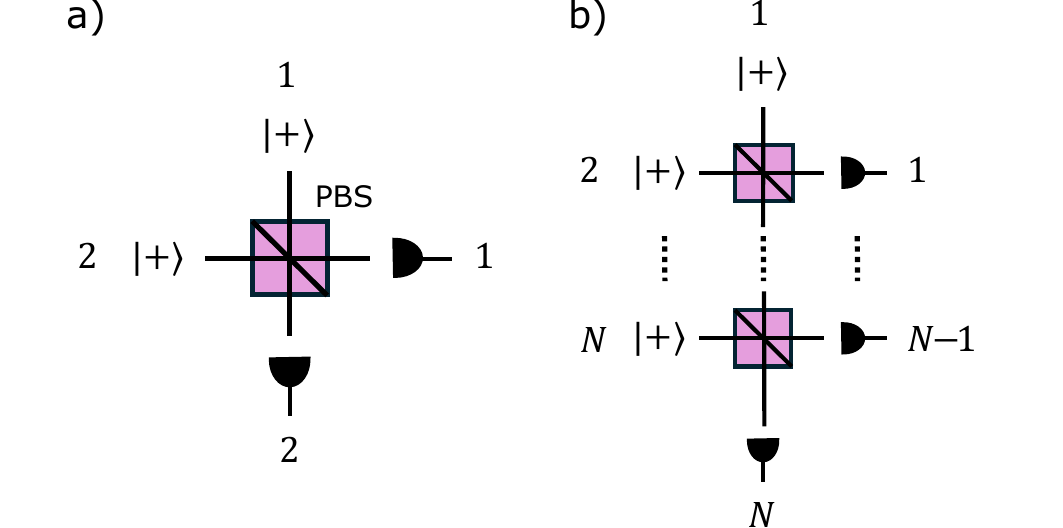}
    \caption{\justifying Postselection scheme to generate polarization-encoded $N$-photon GHZ states used as a building block in our photon-weaving quantum server. a) Inserting two photons, in the polarization state $\ket{+}$ each, into a single polarizing beam splitter (PBS) and detecting for coincidences in the two output modes allows for postselecting the two-photon Bell state. b) Inserting $N$ photons, each in the polarization state $\ket{+}$, into $(N-1)$ PBSs and detecting for coincidences in the $N$ output modes allows for postselecting the $N$-photon GHZ state.}
    \label{fig:GHZPostselection}
\end{figure}

Let us fix polarization degrees of freedom, i.e., horizontal ($\ket{H}=\ket{0}$) and vertical ($\ket{V}=\ket{1}$) polarization states.
One of the standard methods for postselecting a GHZ state utilizes $N$ independent photons, each prepared in the state $\ket{+}=(\ket{H}+\ket{V})/\sqrt{2}$, and $(N-1)$ polarizing beam splitters (PBSs), which transmit $\ket{H}$ and reflect $\ket{V}$ (see \cref{fig:GHZPostselection})~\cite{Sagi2003}.
Postselecting for $N$-photon coincidences, i.e., exactly one photon per output mode, can only be successful if all photons have the same polarization, either $H$ or $V$. This happens with a probability of $(1/2)^{N-1}$. Thus, for $N=2$, one postselects the two-photon GHZ state, i.e., the Bell state $\ket{\Psi_{\text{Bell}}}=(\ket{HH}+\ket{VV})/\sqrt{2}$ [see \cref{fig:GHZPostselection}a)]. For $N>2$, one postselects the polarization version of the $N$-photon GHZ state given in \eqref{eq:GHZ} [see \cref{fig:GHZPostselection}b)].

\begin{figure}[t]
    \centering
    \includegraphics[width=0.9\linewidth]{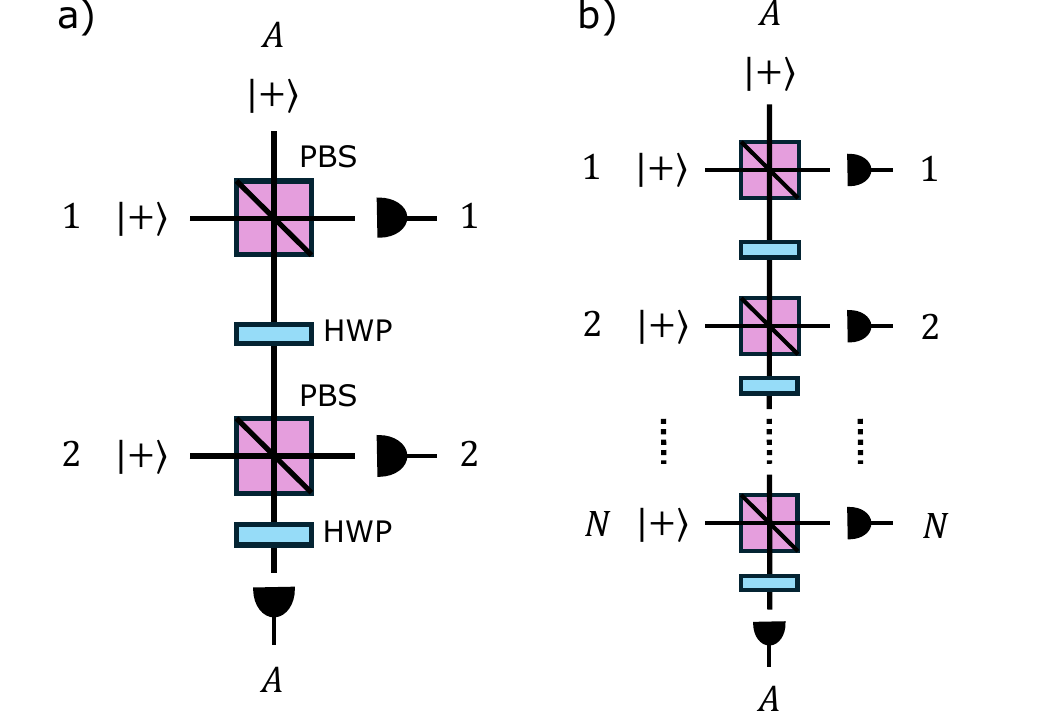}
    \caption{\justifying Postselection scheme to generate polarization-encoded $N$-photon path graph states used as a building block in our photon-weaving quantum server. a) Photonic CZ gate using one auxiliary photon and postselection~\cite{Pittman2003}. Two input photons 1 and 2 plus an auxiliary photon $A$ prepared in the state $\ket{+}$ each interfere at a combination of two polarizing beam splitters (PBSs) and half-wave plates (HWPs). Postselecting for one photon per output mode completes the CZ operation between the two input photons. b) Array of $(N-1)$ photonic CZ gates using a single auxiliary photon to postselect a path graph state~\cite{Pilnyak2017}. Postselecting for one photon per output mode demonstrates the successful execution of $(N-1)$ CZ gates and, therefore, postselects the $N$-photon path graph state.}
    \label{fig:CPhase}
\end{figure}

\subsubsection{Postselecting Path Graph States}

To postselect path graph states, one simply has to add half-wave plates (HWPs) (see \cref{fig:CPhase}) to the GHZ state scheme (see \cref{fig:GHZPostselection})~\cite{Pittman2003,Pilnyak2017}.
The reason for this is that HWPs at an angle of $22.5^{\circ}$ rotate the polarization states $\ket{H/V}$ to $\ket{+/-}$. Note that HWPs at an angle of $0^{\circ}$ perform Pauli $Z$ gates, which, up to a phase, allows for postselecting GHZ states as described above. Hence, one can easily switch between the two schemes by simply rotating the HWPs. Throughout this paper, whenever mentioning HWPs we will assume an angle of $22.5^{\circ}$.

In the case of three photons [see ~\cref{fig:CPhase} a)], this scheme describes a photonic CZ gate~\cite{Pittman2003}, where the initial state is given by
\begin{align}
\label{eq:InitialState}
    \ket{\Psi} = \ket{+}_{A}\ket{+}_{1}\ket{+}_{2} .
\end{align}
Successful action of the first PBS and the HWP changes this state to
\begin{align}
\label{eq:Psi'}
   & \xrightarrow{\text{PBS}} \frac{1}{2} \big( \ket{H}_{A}\ket{H}_{1} + \ket{V}_{A}\ket{V}_{1} \big)\ket{+}_{2} \nonumber \\
   & \xrightarrow{\text{HWP}} \frac{1}{2} \big( \ket{+}_{A}\ket{H}_{1} + \ket{-}_{A}\ket{V}_{1} \big)\ket{+}_{2}.
\end{align}
Again, we are postselecting for $N$-photon coincidences and have left out the terms not fulfilling the postselection criterion. Due to this, the state is now normalized by the success probability, which is $1/2$. The state after the second set of PBS and HWP takes the postselected form
\begin{align}
\label{eq:3LinearGraph}
    & \xrightarrow{\text{PBS}} \frac{1}{4} \big( \ket{H}_{A}\ket{H}_{1}\ket{H}_{2} + \ket{V}_{A}\ket{H}_{1}\ket{V}_{2}   \nonumber \\
    & \phantom{\xrightarrow{\text{HWP}}{}} + \ket{H}_{A}\ket{V}_{1}\ket{H}_{2} - \ket{V}_{A}\ket{V}_{1}\ket{V}_{2} \big)\nonumber \\
   & \xrightarrow{\text{HWP}} \frac{1}{4} \big( \ket{+}_{A}\ket{H}_{1}\ket{H}_{2} + \ket{-}_{A}\ket{H}_{1}\ket{V}_{2}   \nonumber \\
    & \phantom{\xrightarrow{\text{HWP}}{}}  + \ket{+}_{A}\ket{V}_{1}\ket{H}_{2} - \ket{-}_{A}\ket{V}_{1}\ket{V}_{2} \big) .
\end{align}
From this, one can see that the phase change $+1 \rightarrow -1$ only occurs if photons 1 and 2 are both in the state $\ket{V}$.

Following the second HWP, the state given in lines 3 and 4 of \eqref{eq:3LinearGraph} is identical to a three-qubit path graph state. This means that measuring out the auxiliary qubit in the $H/V$ basis projects the other two qubits into the two-qubit path graph state and completes the CZ operation. The success probability is $1/2\times 1/2 =1/4$.

Note that the scheme can also be interpreted as two sequential type-I fusion gates~\cite{Browne2005}. The photon detected in the standard type-I fusion is recycled and used again for the second type-I fusion gate.

Independently of the chosen interpretation, one can see that the performed operation can be continued, i.e., the auxiliary photon can interfere with more photons and thereby implement more CZ/type-I fusion gates [see \cref{fig:CPhase} b)]. Since each pair of subsequent photons in the identical state $\ket{V}$ leads to a sign change, in total, this procedure postselects an $N$-qubit path graph state ($(N+1)$ when including the auxiliary photon) with a probability of $(1/2)^{N}$~\cite{Pilnyak2017,Meyer-Scott2022}.

\subsubsection{Postselected Photon Weaving}

\begin{figure}
    \centering
    \includegraphics[width=0.9\linewidth]{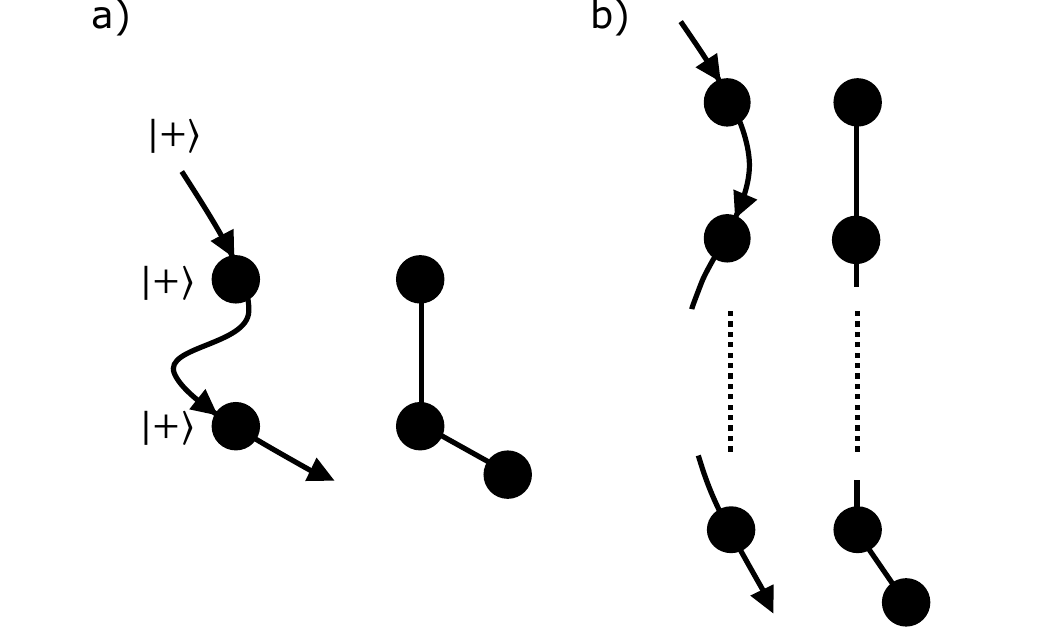}
    \caption{\justifying Photon-weaving a path graph state with a single auxiliary weaving photon (see \cref{fig:CPhase})~\cite{Lin2011}. a) Weaving two photons results in a two-photon graph state with the weaving photon connected to the second photon only, i.e., in total, a three-photon path graph state. b) Weaving $N$ photons results in an $N$-photon graph state with the weaving photon connected to the $N$th photon only, i.e., in total, an $(N+1)$-photon path graph state.}
    \label{fig:LinearSewing}
\end{figure}

The scheme for postselecting path graph states has also been introduced in a more general form as \textit{photon weaving}~\cite{Lin2011}. As can be seen in \cref{fig:LinearSewing}, the auxiliary photon weaves together the photons on its way, i.e., it entangles them through CZ operations.
Note that in this work, besides the term \textit{(graph-state) weaving}, we will also use the term \textit{GHZ-state weaving} depending on the weaving mechanism used. This means that using only PBSs as described first corresponds to GHZ-state weaving while using PBSs and HWPs as described second corresponds to graph-state weaving or simply weaving.

\begin{figure}[t]
    \centering
    \includegraphics[width=0.9\linewidth]{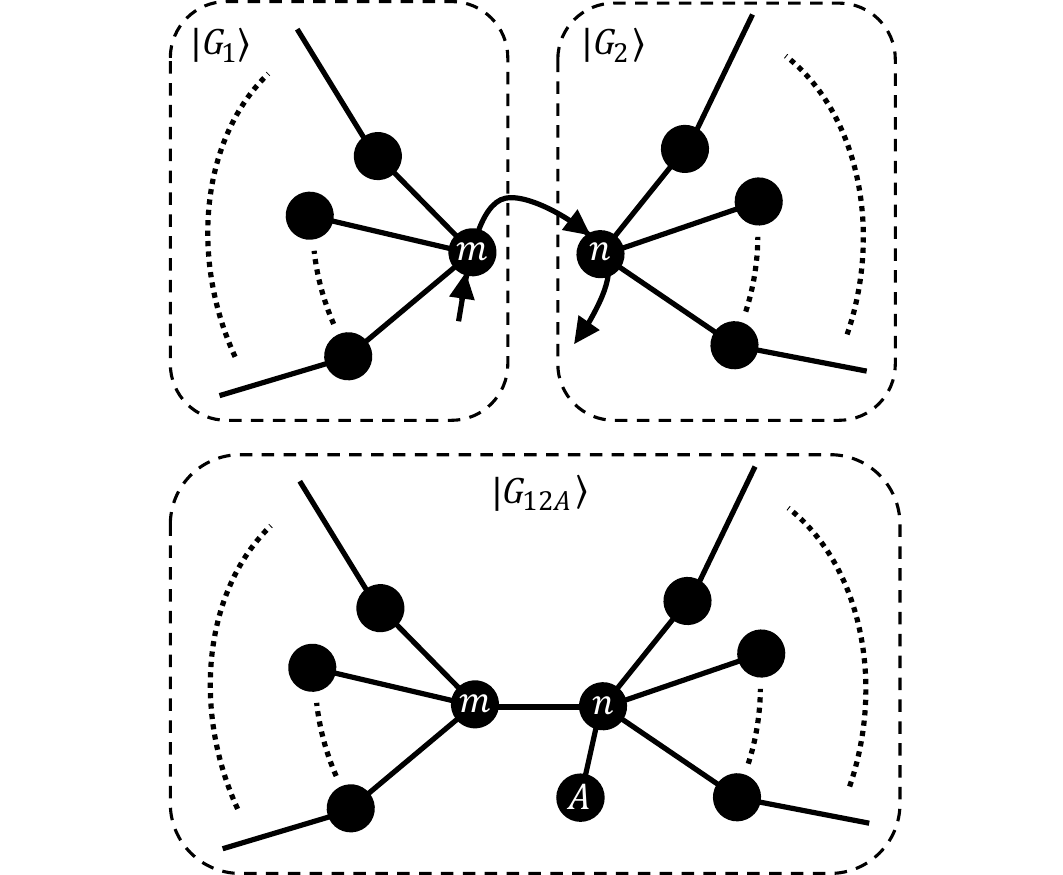}
    \caption{\justifying Fusing two separate graph states $\ket{G_{1}}$ and $\ket{G_{2}}$ by photon-weaving their respective qubits $m$ and $n$ with a single weaving photon $A$ into the state $\ket{G_{12A}}$.}
    \label{fig:SewingGraphs}
\end{figure}

In general, graph-state weaving allows two or more graph states to be weaved together~\cite{Lin2011}, i.e., in effect, applying CZ gates between them.
To understand this, we start by looking at two arbitrary graph states $\ket{G_{1}}$ and $\ket{G_{2}}$. W.l.o.g., we assume that we want to weave $\ket{G_{1}}$ and $\ket{G_{2}}$ together using their qubits $m$ and $n$, respectively. We know that we can rewrite the two states as
\begin{align}
    \ket{G_{1/2}} = \frac{1}{\sqrt{2}} \left( \ket{g_{1/2,H}}\ket{H}_{m/n} + \ket{g_{1/2,V}}\ket{V}_{m/n} \right) ,
\end{align}
with
\begin{align}
    \ket{g_{1/2,H/V}} = \sqrt{2} \bra{H/V}_{m/n} \ket{G_{1/2}} .
\end{align}
The complete state, including the weaving photon before entering the photonic CZ gate, takes the form
\begin{align}
\label{eq:G1G2A}
    \ket{+}_{A}\ket{G_{1}}\ket{G_{2}} & = \frac{1}{2} \big( \ket{+}_{A}\ket{g_{1,H}}\ket{H}_{m}\ket{g_{2,H}}\ket{H}_{n}  \nonumber \\
    & \phantom{={}} + \ket{+}_{A}\ket{g_{1,H}}\ket{H}_{m}\ket{g_{2,V}}\ket{V}_{n} \nonumber \\
    & \phantom{={}} + \ket{+}_{A}\ket{g_{1,V}}\ket{V}_{m}\ket{g_{2,H}}\ket{H}_{n} \nonumber \\
    & \phantom{={}}  + \ket{+}_{A}\ket{g_{1,V}}\ket{V}_{m}\ket{g_{2,V}}\ket{V}_{n} \big) .
\end{align}
Comparing \eqref{eq:G1G2A} to \eqref{eq:InitialState}-\eqref{eq:3LinearGraph}, we can directly see the action of the CZ gate, which leads to
\begin{align}
\label{eq:G1G2ACPhase}
    \ket{G_{12A}} & = \frac{1}{4} \left( \ket{+}_{A}\ket{g_{1,H}}\ket{H}_{m}\ket{g_{2,H}}\ket{H}_{n} \right. \nonumber \\
    & \phantom{={}} + \ket{-}_{A}\ket{g_{1,H}}\ket{H}_{m}\ket{g_{2,V}}\ket{V}_{n} \nonumber \\
    & \phantom{={}} + \ket{+}_{A}\ket{g_{1,V}}\ket{V}_{m}\ket{g_{2,H}}\ket{H}_{n} \nonumber \\
    & \phantom{={}} \left. - \ket{-}_{A}\ket{g_{1,V}}\ket{V}_{m}\ket{g_{2,V}}\ket{V}_{n} \right) ,
\end{align}
where the qubits $m$ and $n$ of $\ket{G_{1}}$ and $\ket{G_{2}}$, respectively, are now connected (see \cref{fig:SewingGraphs}). Additionally, the auxiliary qubit is solely connected to qubit $n$. It can either be removed by performing a $Z$ measurement, i.e., a measurement in the $H/V$ basis, or used to connect the next graph state. The postselected gate connecting two graph states succeeds with probability $1/4$. In total, weaving together $N$ graph states with the help of a weaving photon succeeds with probability $(1/2)^{N}$. Again, this can be interpreted as subsequent type-I fusion operations, where one of the photons is being reused for the next fusion operation.

Due to the postselection, the weaving photon cannot interact with the same photon twice. Hence, to weave a \textit{cycle graph state} postselectively, the weaving photon must be part of an initial two-photon graph state of the form $\ket{G_{\text{Bell}}}=(\ket{+H}+\ket{-V})/\sqrt{2}$, which is equivalent to a single-photon rotated Bell state.
Weaving then gives a path graph state that can be closed by weaving, i.e., fusing the first and the last photon. The postselected result is a cycle graph state with an additional photon connected to the circle (see \cref{fig:CircularWeaving}; see Appendix~\ref{app:WeavingCircularGraphState} for mathematical proof).

\begin{figure}
    \centering
    \includegraphics[width=0.9\linewidth]{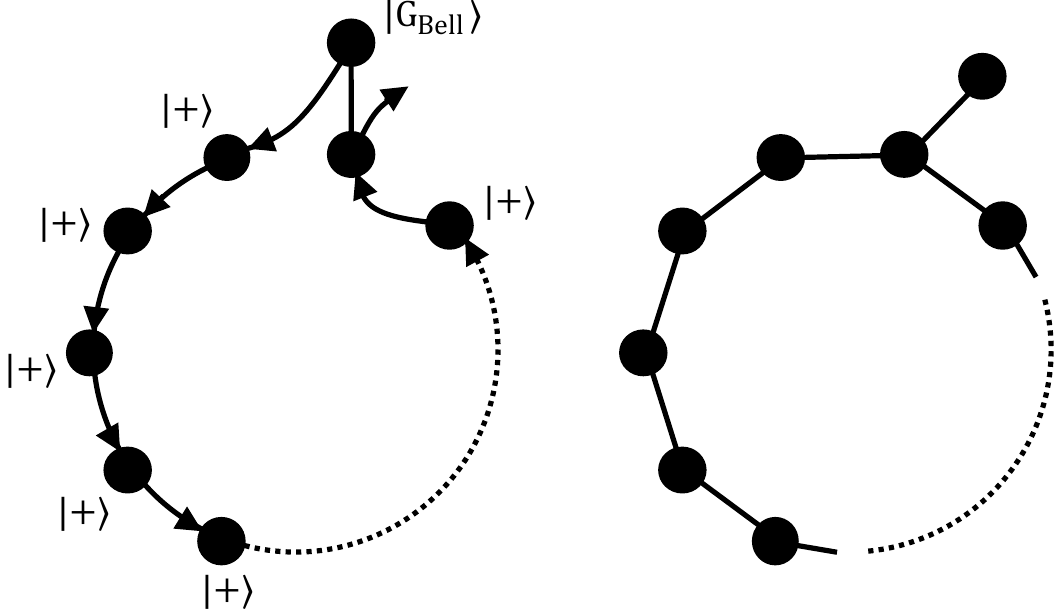}
    \caption{\justifying Photon-weaving a cycle graph state with a weaving photon belonging to an auxiliary two-photon graph state $\ket{G_{\text{Bell}}}$. First, a path graph state is weaved, with the two auxiliary photons being the first and last qubits of the path graph. Second, the path graph is closed to a cycle graph state by additionally weaving the outer photons. The weaving photon is connected to the final cycle graph state but is not included.}
    \label{fig:CircularWeaving}
\end{figure}

Finally, combining GHZ weaving and path-graph or cycle-graph weaving generates caterpillar graphs~\cite{bui2013note, coiteux2024genuinely, huet2024deterministic}. Here, caterpillar graphs are path graph states with additional leaves, which can be thought of as a connected chain of GHZ states. Similarly, this has been discussed for deterministic single photon emitters, combining the emission of photonic GHZ and photonic path graph states~\cite{Hilaire2023}.

\section{Photon-Weaving Quantum Server}

In this section, we introduce our PWQS that acts as a central node and, simultaneously, as a possible user of a quantum network.
It can distribute different types of entangled states to $M$ users by photon weaving, namely GHZ states, path graph states, cycle graph states, and caterpillar graph states (see Appendix~\ref{sec:zigzag_vertex_minors} for a derivation and complete classification of states that can be extracted).
For this, the PWQS must consist of PBSs, HWPs, Bell states, and single-photon detectors. Furthermore, each user must share an entangled two-photon graph state with the PWQS.

First, we describe a PWQS using linear optics only, where all photons must be simultaneously weaved. Second, we describe an adjusted PWQS, including the possibility of storing photons. It weaves smaller entangled states first, which are then fused together.
Note that neither of the two versions takes into account the experimental implementation of the qubits belonging to the users. Depending on the chosen design, these could be photons or long-lived memories. We simply assume that the users can store their qubits and perform any required single-qubit operations.

\subsection{Without Storing Photons}\label{sec:no_storage}

\begin{figure}
    \centering
    \includegraphics[width=0.9\linewidth]{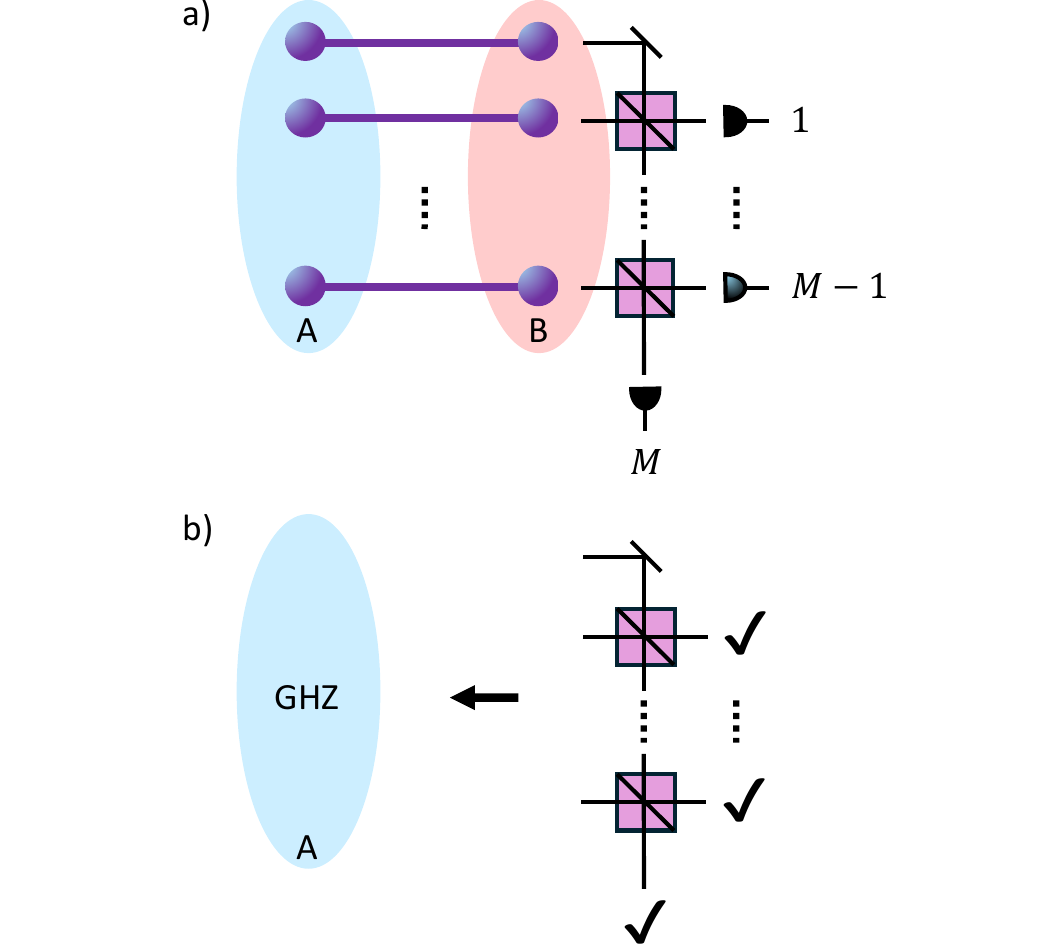}
    \caption{\justifying Photon-weaving quantum server (PWQS) protocol to distribute GHZ states to $M$ users~\cite{Wang2009}. a) $M$ photons belonging to the $M$ Bell pairs shared between the users (A) and the PWQS (B) are sent into $(M-1)$ PBSs for GHZ-state weaving. b) Detecting $M$-photon coincidences completes the GHZ-weaving and heralds the successful distribution of a GHZ state to the users.}
    \label{fig:GHZSwapping}
\end{figure}

\textbf{GHZ state protocol.} The first protocol the PWQS can perform is the standard scheme to generate $M$-user GHZ states from $M$ polarization-encoded two-photon graph states (see \cref{fig:GHZSwapping})~\cite{Wang2009}. Each of the shared two-photon graph states is thereby prepared in the state $\ket{G_{\text{Bell}}}=(\ket{+H}+\ket{-V})/\sqrt{2}$. Identically to the GHZ-state weaving discussed in \cref{sec:background} (see \cref{fig:GHZPostselection}), the $M$ photons reaching the PWQS are sent into $(M-1)$ PBSs. Coincident detection of $M$ photons then verifies that all $M$ photons have had the same polarization. Detecting in the $\pm$-basis further guarantees that the specific polarization remains unknown and the $M$ photons shared with the users are projected into a GHZ state of the form~\cite{Wang2009}
\begin{align}
    \ket{\text{GHZ}_{M}} = \frac{1}{\sqrt{2}} \left( \ket{++\ldots +} + (-1)^{M_{-}}\ket{--\ldots -} \right),    
\end{align}
where $M_{-}$ denotes the number of photons measured in the state $\ket{-}$.
Hereby, the PWQS can freely decide if it wants to participate in the communication or not. To participate, it will employ one of the coincidentally detected photons to become part of the shared GHZ state. Assuming no loss, the success probability is given by $p_{\text{GHZ}}=(1/2)^{M-1}$.

\begin{figure}
    \centering
    \includegraphics[width=0.9\linewidth]{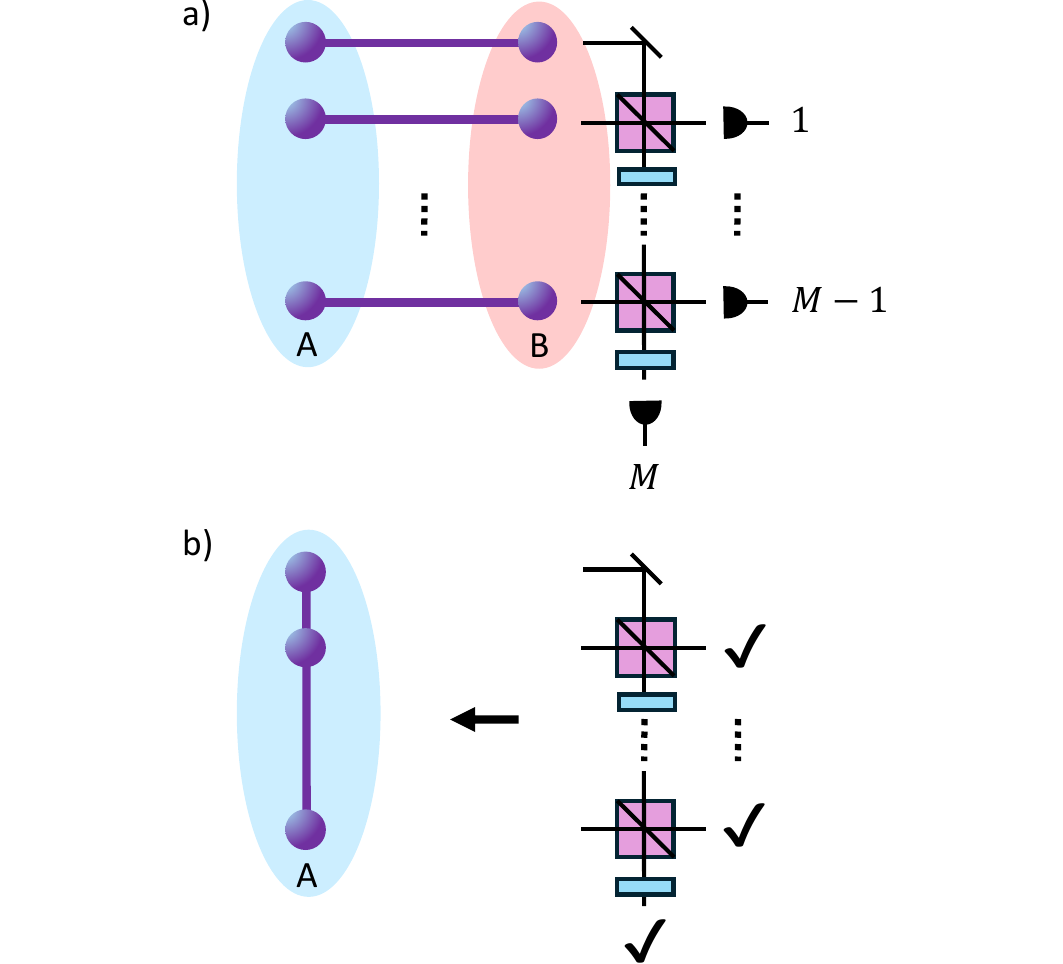}
    \caption{\justifying Photon-weaving quantum server (PWQS) protocol to distribute path graph states to $M$ users. a) $M$ photons belonging to the $M$ two-photon graph states shared between the users (A) and the PWQS (B) are sent into $(M-1)$ PBSs for weaving a path graph state. b) Detecting $M$-photon coincidences completes the graph-weaving and heralds the successful distribution of a path graph state to the users.}
    \label{fig:LinearSwapping}
\end{figure}

\textbf{Path graph state protocol.} The second possible protocol is based on photon weaving and a sequence of $X$ measurements~\cite{Hahn2019}, i.e., measurements in the $\pm$ basis, to generate an $M$-photon path graph state connecting the users (see \cref{fig:LinearSwapping}). Again, the PWQS can decide if it wants to participate in the generated graph state or not.

\begin{figure}
    \centering
    \includegraphics[width=0.9\linewidth]{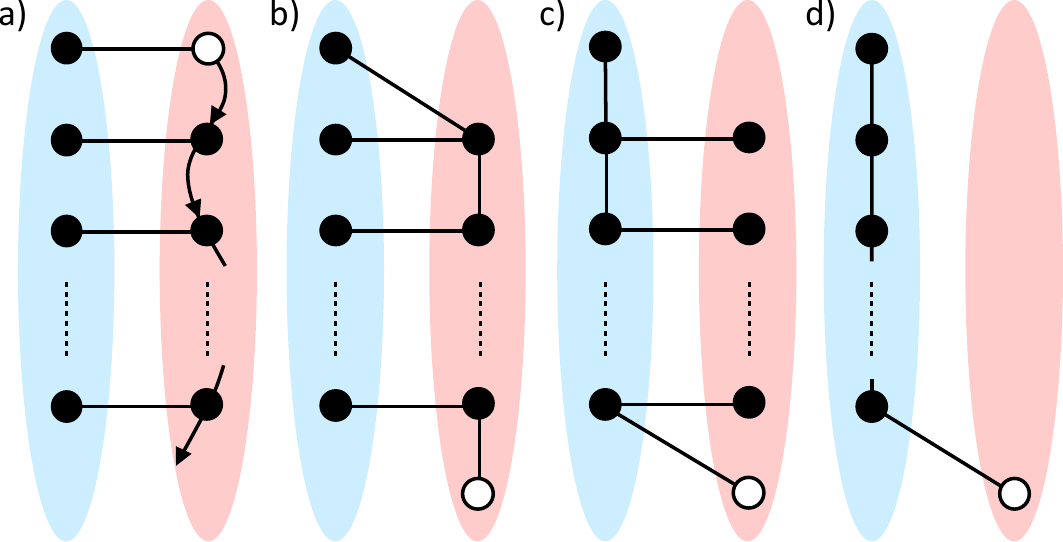}
    \caption{\justifying Scheme to distribute path graph states by graph-state weaving. a) Photon-weaving a graph state using one of the input photons as a weaving photon yields b) a comb-like graph state. Performing measurements in the $X$-basis c) flips the comb and d) generates the path graph state, including the quantum server.}
    \label{fig:WeavingComb}
\end{figure}

Using one of the shared photons as a weaving photon generates the comb-like graph state as shown in \cref{fig:WeavingComb}. Subsequently, performing $X$ measurements transforms this state into a path graph state in which the PWQS possesses one of the outer qubits. Under the assumption that all $M$ photons have arrived simultaneously, the success probability equals $p_{\text{path}}=(1/2)^{M-1}$.

As discussed in \cite{Hilaire2023}, we can interpret the leaf qubits/users in the comb-like graph as redundantly encoded qubits.
User and PWQS qubits can change roles through Hadamard rotations, i.e., basis changes. This means that measuring the PWQS qubits in the $X$-basis projects the users into a path graph state, where all users except the first one would need to perform Hadamard rotations. Additionally, each PWQS qubit measured in the state $\ket{-}$ requires a $Z$ operation of its connected user. Therefore, local unitary corrections are only required from the users and can all be moved to the post-processing. Hence, the PWQS photons can be measured simultaneously without any additional corrections needed.

\begin{figure}
    \centering
    \includegraphics[width=0.9\linewidth]{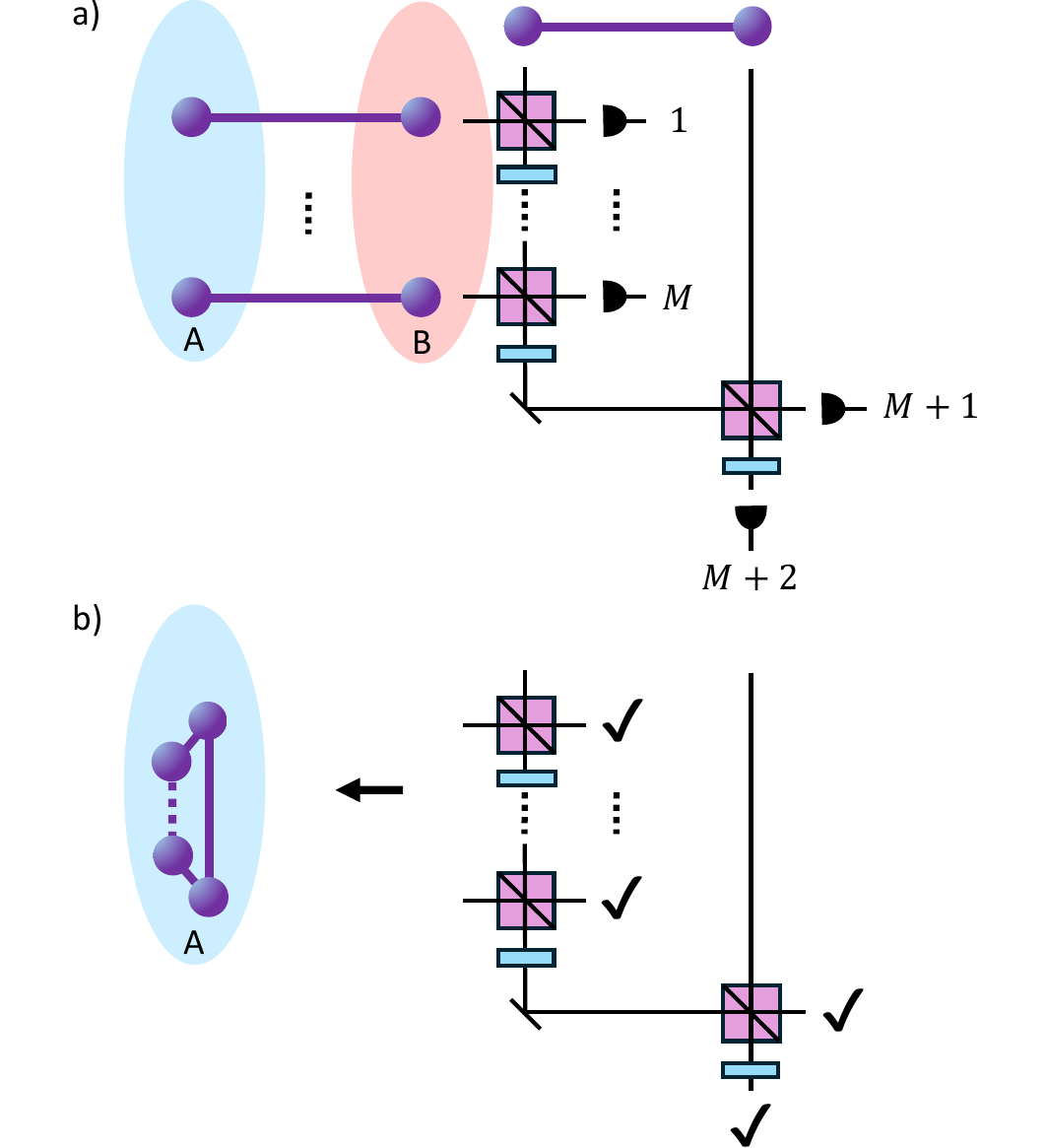}
    \caption{\justifying Photon-weaving quantum server (PWQS) protocol to distribute cycle graph states to $M$ users. a) $M$ photons belonging to the $M$ two-photon graph states shared between the users (A) and the PWQS (B) plus a weaving photon belonging to a two-photon graph state owned by the PWQS are sent into $(M+1)$ PBSs for weaving a cycle graph state. b) Detecting $M+2$ coincidences completes the graph-weaving and heralds the successful distribution of a cycle graph state to the users.}
    \label{fig:CircularSwapping}
\end{figure}

\begin{figure}
    \centering
    \includegraphics[width=0.9\linewidth]{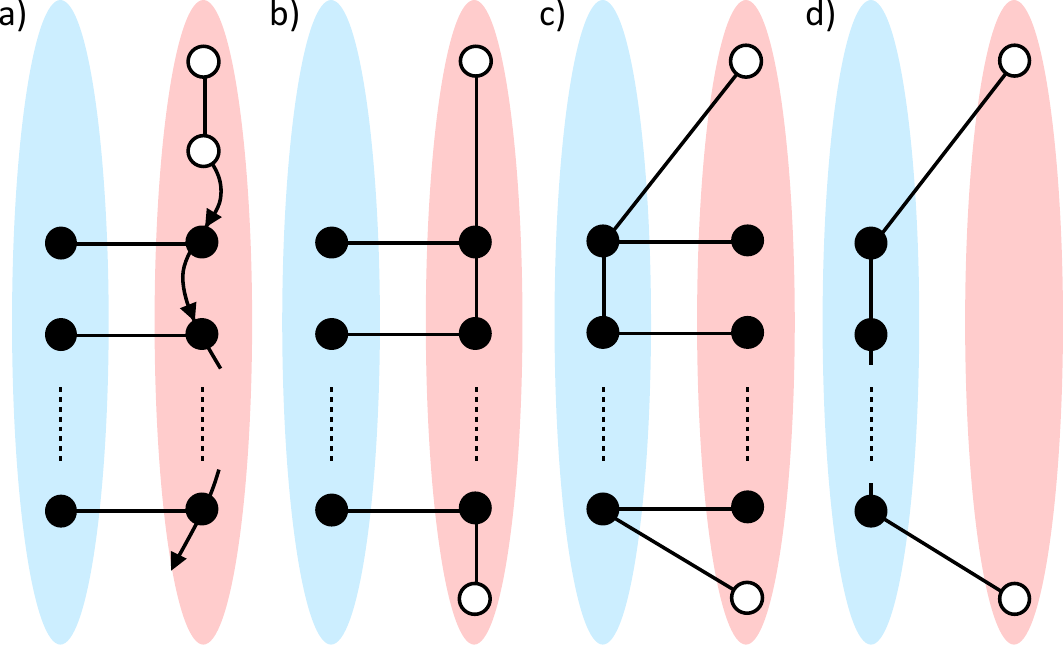}
    \caption{\justifying Scheme to distribute cycle graph states by graph-state weaving. a) Photon-weaving a graph state using a single weaving photon belonging to a two-qubit graph state owned by the photon-weaving quantum server (PWQS) yields b) a comb-like graph state. Performing measurements in the $X$-basis c) flips the comb and d) generates the path graph state, where the PWQS owns the two outer photons. Finally, fusing the two outer photons distributes a cycle graph state (see \cref{fig:CircularWeaving}).}
    \label{fig:WeavingCircle}
\end{figure}

\textbf{Cycle graph state protocol.} The third protocol is similar to generating a path graph state, except that the weaving photon is now part of a Bell pair that belongs to the PWQS completely (see \cref{fig:CircularSwapping,fig:WeavingCircle}). Performing $X$ measurements transforms this state into a path graph state where the PWQS possesses both the first and the last qubit. Finally, fusing the two outer qubits, both at the PWQS, generates a cycle graph state that includes the PWQS. As before, assuming no losses, the success probability is given by $p_{\text{cycle}}=(1/2)^{M+1}$.

\textbf{Caterpillar graph state protocol.} By combining the different methods, one can distribute caterpillar graph states to the users. The success probability of generating a caterpillar graph is $p_{\text{pathCP}}=(1/2)^{M-1}$. Weaving with an additional Bell pair to close the caterpillar and generate a `cycle caterpillar graph' has a success probability of $p_{\text{cycleCP}}=(1/2)^{M+1}$.

\subsection{With Storing Photons}\label{sec:storage}

Due to the exponential decrease of the success probabilities and high photon losses, schemes completely based on postselective photon weaving, as described above, can be impractical for increasing numbers of users. In this section, we show that our PWQS can also be adjusted to become practical in this regime. By introducing the ability to store photons, we present a PWQS that first weaves four-photon building blocks and then fuses them into larger graph states step by step. At the same time, unsuccessful fusion operations do not let the complete weaving process fail.

\begin{figure}
    \centering
    \includegraphics[width=0.9\linewidth]{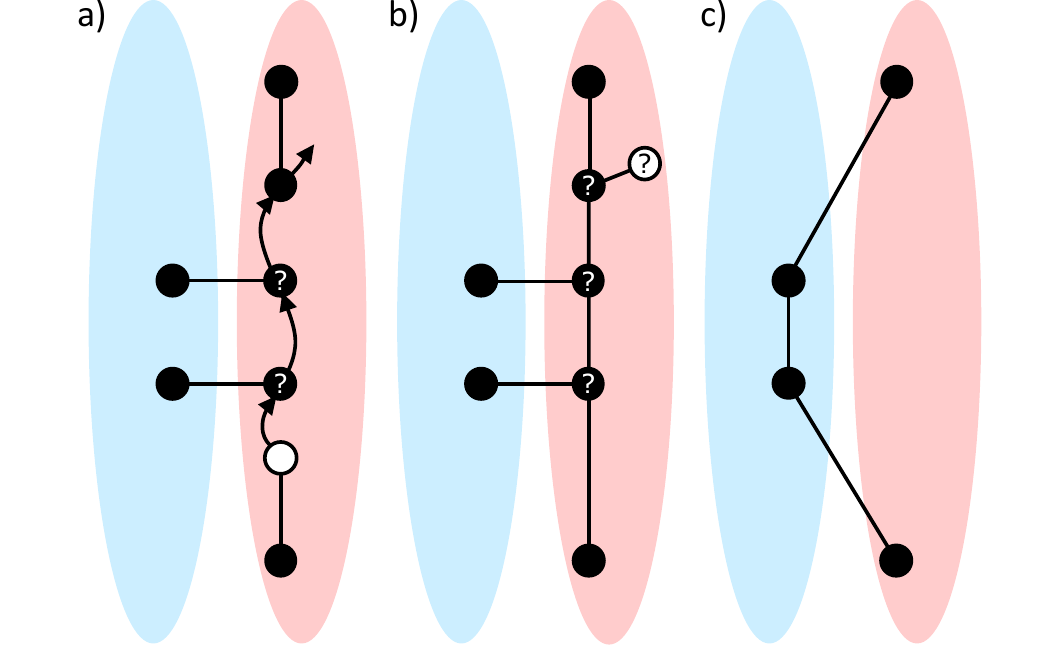}
    \caption{\justifying Scheme to weaving two users to generate a path-shaped four-photon building block. a) Photon-weaving a graph state using a single weaving photon belonging to a two-qubit graph state owned by the photon-weaving quantum server (PWQS) yields b) a comb-like graph state. Performing measurements in the X-basis c) generates the path-shaped four-photon building block, where the PWQS owns the two outer photons. Note that due to noise, the PWQS can not be sure the photons shared with the users have arrived until all photons involved in the weaving have been measured. Nodes affected by this uncertainty are marked with question marks.}
    \label{fig:BuildingBlock}
\end{figure}

To generate a path-shaped four-photon building block, the PWQS weaves two photons belonging to Bell pairs shared with two users. As shown in \cref{fig:BuildingBlock}, the weaving photon belongs to a deterministic Bell pair solely owned by the PWQS. In addition to the two shared Bell pairs, a second Bell pair owned by the PWQS is connected by weaving. Similar to the discussion above, this generates a comb-like graph state. Finally, the PWQS measures all photons involved in the weaving process to ensure that the shared photons have arrived and the weaving was successful. This generates the path-shaped four-photon building block in which the PWQS owns and stores the two outer photons (see \cref{fig:BuildingBlock}).

Instead of graph-state weaving, the PWQS could also perform GHZ weaving. Measuring all photons involved in the weaving process generates a star-shaped four-photon building block. Assuming no photon loss, the success probability for generating a four-photon building block, either path or star-shaped, is upper bounded by $p_{\text{BB}}=1/8$.

\begin{figure}[b]
    \centering
    \includegraphics[width=1\linewidth]{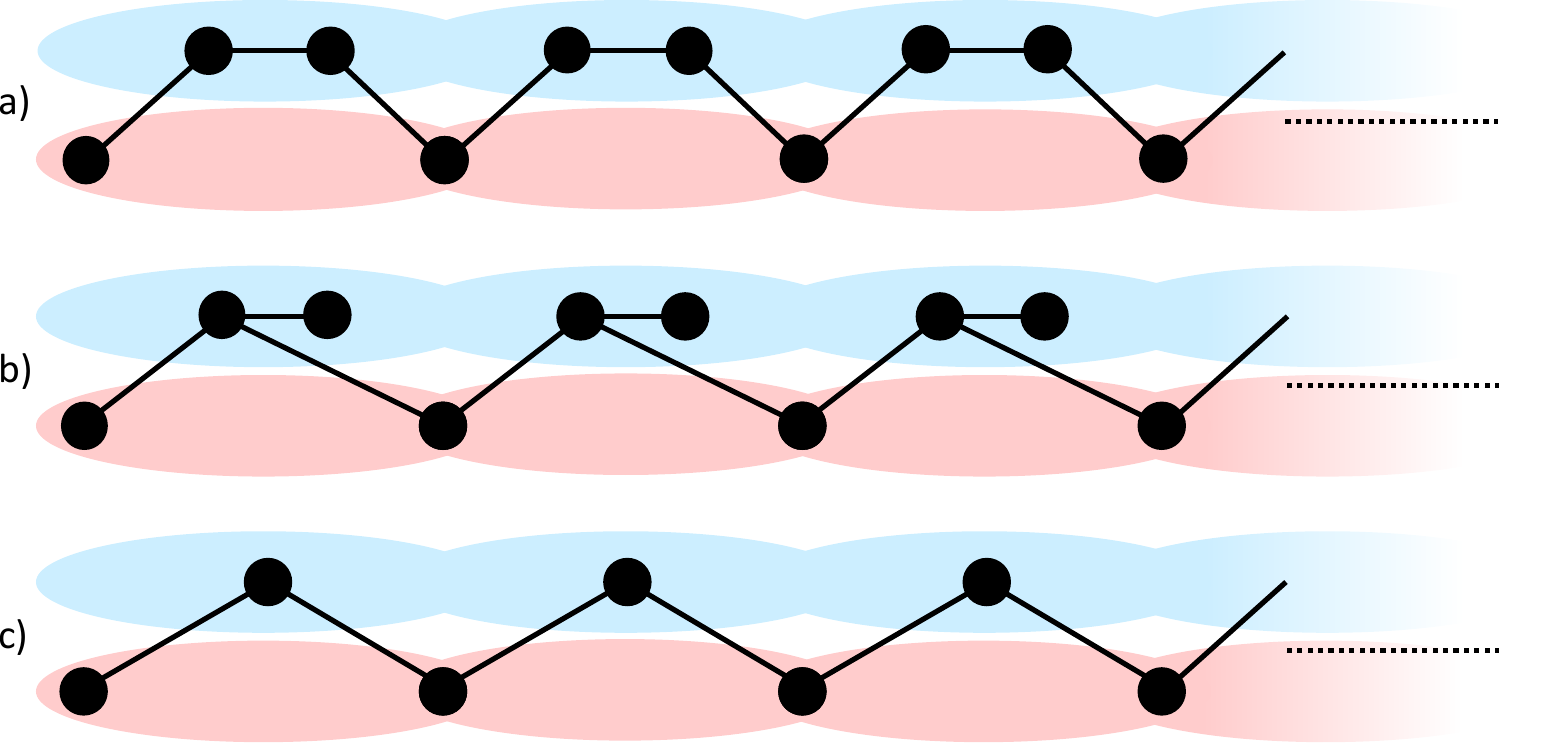}
    \caption{\justifying Different graph states that can be shared between the photon-weaving quantum server and the users from fusing different building blocks. a) Path graph state from fusing path-shaped four-photon building blocks. b) Partial honeycomb from fusing star-shaped four-photon building blocks. c) Zigzag graph state from fusing three-photon building blocks.}
    \label{fig:FusingBuildingBlocks}
\end{figure}

Now, to generate an $M$-user graph state, the PWQS fuses multiple building blocks (see \cref{fig:FusingBuildingBlocks}). Hereby, each fusion measures one photon~\cite{Browne2005}, and the second is stored at least until the next fusion is performed. In general, any fusion operation might fail with a probability of $1/2$. However, a failed fusion does not mean all steps must be repeated. In particular, only the last two users must be removed from the graph by $Z$ measurements and entangled with the PWQS anew. This means that only the last building block must be repeated.

Instead of working with four-photon building blocks connecting two users, the PWQS could also generate three-photon building blocks connecting only one user. Although the success probability per building block is increased to $p_{\text{BB}}=1/4$, it also requires twice as many Bell pairs generated in the PWQS. Fusing such three-photon building blocks will generate a path graph state in a zigzag arrangement between users and the PWQS (similar as discussed in \cite{Mazza2025}).

Finally, depending on the used building blocks and their resulting graph states (see \cref{fig:FusingBuildingBlocks}), the PWQS can now distribute different states to the users by choosing different measurements.
For example, path-shaped four-photon building blocks and $Y$ measurements distribute a path graph state, while star-shaped four-photon building blocks and $X$ measurements distribute a GHZ state.
Similarly, three-photon building blocks and either $X$ or $Y$ measurements distribute a path graph state or a GHZ state, respectively.
Note that combining different types of building blocks allows for the distribution of additional graph states. For example, combining path and star-shaped building blocks distributes caterpillar graph states. Note further that fusing the two outer photons in the PWQS creates cycle graph states.
We provide a full characterization of all states that can be distributed to the users in this fashion in Appendix~\ref{sec:zigzag_vertex_minors}.

\section{Conclusion}
\label{sec:Conclusion}

In this paper, we introduced a flexible so-called photon-weaving quantum server (PWQS) based on linear optical elements only that can distribute different locally nonequivalent graph states.
Our PWQS is based on two different photonic fusion protocols, i.e., GHZ-state weaving and graph-state weaving.
We showed that depending on the chosen protocol, our PWQS can distribute GHZ states, as well as, path, cycle, and caterpillar graph states.

We discussed two different possibilities for implementing our PWQS: 
i) Without the possibility of storing photons, i.e., all photons have to be fused and measured simultaneously. This version of the PWQS allows for a simple and experimentally cheap distribution of different graph states to a network but with an exponentially decreasing success probability.
ii) With the possibility of storing photons, which weaves small building blocks connecting one or two users first. The building blocks are only fused in a second step and can be prepared in parallel. Further, the failure of a single fusion of two building blocks does not make the complete fusion process fail.

For the sake of simplicity, we focused on linearly weaving graph states and fusing building blocks. The only exception discussed is cycle graph states. However, it will be interesting to investigate schemes that weave graph states and fuse building blocks in a two-dimensional manner, e.g., see \cite{Mazza2025} for a discussion of two-dimensionally fusing zigzag graph states.

Another helpful extension of our scheme would be allowing the PWQS to decide which users participate and in which order to weave them. This could be done by utilizing programmable photonic circuits~\cite{Bogaerts2020}. In combination with considering a network of multiple PWQSs, this would allow for distributing larger and more diverse graph states.

Other future directions might be to adjust the photon-weaving scheme to a heralding scheme without a central quantum server, as it has been discussed in \cite{Zo2025} for GHZ states, or extend photon weaving to higher dimensions~\cite{Luo2019}.

Finally, it will be important to discuss our PWQS in the presence of losses and the necessary overhead. This will be necessary to compare it to other schemes under realistic conditions~\cite{Bugalho2023,avis2023analysis}.

\section*{Acknowledgment}

We thank Maria Flors Mor-Ruiz and Jorge Miguel-Ramiro for their helpful comments on the manuscript.

\appendices

\section{Weaving Cycle Graph State: Mathematical Proof}
\label{app:WeavingCircularGraphState}

An arbitrary graph state $\ket{G_{f,l}}$, for example, a path graph state, is given. We briefly discuss fusing two specific qubits $f$ and $l$ of the graph, letting the two photons interfere at a PBS and postselecting for coincidences. The graph state can be rewritten in the following form:
\begin{align}
\label{eq:Gfl}
    \ket{G_{f,l}} &= \frac{1}{2} \left( \ket{g_{H,H}}\ket{H}_{f}\ket{H}_{l} + \ket{g_{H,V}}\ket{H}_{f}\ket{V}_{l} \right. \nonumber \\
    & \phantom{={}} \left. + \ket{g_{V,H}}\ket{V}_{f}\ket{H}_{l} + \ket{g_{V,V}}\ket{V}_{f}\ket{V}_{l} \right) ,
\end{align}
with
\begin{align}
    \ket{g_{H/V,H/V}} = 2 \bra{H/V}_{f}\bra{H/V}_{l} \ket{G_{f,l}} .
\end{align}
Detecting coincidences after the PBS now postselects the two terms of \eqref{eq:Gfl} in which qubit $f$ and qubit $l$ are in the same state, i.e.,
\begin{align}
\label{eq:Gfl'}
    & \xrightarrow{\text{PBS}} \frac{1}{2} \left( \ket{g_{H,H}}\ket{H}_{f}\ket{H}_{l}  + \ket{g_{V,V}}\ket{V}_{f}\ket{V}_{l} \right) \nonumber \\
   & \xrightarrow{\text{HWP}} \frac{1}{2} \left( \ket{g_{H,H}}\ket{H}_{f}\ket{+}_{l}  + \ket{g_{V,V}}\ket{V}_{f}\ket{-}_{l} \right) ,
\end{align}
where we additionally have rotated the state of qubit $l$ by use of a HWP to complete the graph-state fusion. From \eqref{eq:Gfl'} we can see that all qubits previously connected to either qubit $f$ or qubit $l$ are now solely connected to qubit $f$. Further, qubit $l$ is only connected to qubit $f$. This proves the graph state's form after fusing two of its own qubits. In particular, it proves that a path graph state is fused into a cycle graph state with an additional qubit attached (see \cref{fig:CircularWeaving}).

\section{The Class of Extractable States}\label{sec:zigzag_vertex_minors}

Our goal is to characterize the possible resulting states obtained by starting with a zigzag state and selectively measuring certain qubits (see Fig.~\ref{fig:FusingBuildingBlocks}c).

First, we note that the analysis of the zigzag state also extends to the partial honeycomb (see Fig.~\ref{fig:FusingBuildingBlocks}b); this is because we only measure the qubits that are not connected to a leaf qubit. As such, performing these measurements and applying the CZ gate according to these edges commute. This means we can study the states that we end up when measuring select qubits from a cycle graph, perform the CZ gate afterward (and then potentially do local complementations/single-qubit rotations afterward).
Note that at the end of this section, we briefly discuss the path graph shown in Fig.~\ref{fig:FusingBuildingBlocks}a.

Now, one way to compute the resultant graph state after measurements is by using the relations found in~\cite{hein2004multiparty}. Namely, measuring a vertex $v$ in $Z$ corresponds to deleting the vertex $v$ from the graph. Measuring in $Y$ corresponds to first performing a local complementation and then a $Z$ measurement. Finally, measuring in $X$ corresponds to performing a local complementation on $v$, then a local complementation on an arbitrary connected vertex $w$, then a local complementation on $v$ again, and finally, deleting $v$. If a graph $H$ can be reached from $G$ by performing the above three operations (plus local complementations), we say that $H$ is a \emph{vertex-minor} of $G$.

While the above relations allow us to, in principle, find (representatives of) vertex-minors of the cycle graph, it suffers from two complications. First, the fact that the choice of vertex $w$ is arbitrary when measuring $v$ in $X$ complicates manners. Second, the state that one ends up after a measurement is only equivalent to the above-specified graph state \emph{up to local rotations}. In particular, after only performing an $X$-measurement, the resultant state is not yet a graph state. We thus first have to perform single-qubit Clifford rotations to end up with the desired graph state.

Instead, we will exploit the fact that cycle graphs are \emph{circle graphs}, which we will define shortly. For now, it suffices that a circle graph $G$ can be associated to an Eulerian tour on some $4$-regular multigraph $F$ (with the same vertex set as $G$). An Eulerian tour $U$ on a finite graph is a sequence of vertices $U = \left(v_1, v_2, \ldots, v_1\right)$ such that $v_{i}, v_{i+1}$ are adjacent, and each edge $e$ in the graph is of the form $\lbrace{v_{i}, v_{i+1}\rbrace}$. The key point is that \emph{all} Eulerian tours on $F$ correspond to a graph $G'$ that is equivalent up to local complementations to $G$, i.e.~$G$ and $G'$ are \emph{locally equivalent}. Conversely, every graph $G'$ that is locally equivalent to $G$ corresponds to an Eulerian tour on $F$. In other words, a $4$-regular multigraph (along with its set of Eulerian tours) describes the equivalence class of graphs that are circle graphs. Furthermore, vertex-minors of circle graphs are described by so-called \emph{transition-minors} of the associated $4$-regular graph, which simplifies analysis. Finally, we hope that similar approaches will be of use to other researchers in the field. For the above reasons, we will focus on the $4$-regular graph associated with the (equivalence class of) the cycle graph $C_n$.

Let us now properly define circle graphs and their relation to Eulerian tours on $4$-regular multigraphs. For a more thorough treatment, see~\cite{traldi2015transition, brijder2015isotropic}. A famous result by Euler states that any connected graph whose vertices all have even degree is Eulerian, i.e., there exists an Eulerian tour $U$ on that graph. In particular, any $4$-regular multigraph has an Eulerian tour $U$. For given vertices $v, w$ and an Eulerian tour $U$, we say that $v$ and $w$ interlace if they appear alternating as one follows $U$. Given an Eulerian tour $U$ on a $4$-regular multigraph $F$, the interlacement graph of $U$ is defined as the graph with the same vertex set as $F$, and where $v, w$ are connected if and only if $v$ and $w$ are interlaced with respect to $U$. Not all graphs arise as interlacement graphs from Eulerian tours on $4$-regular multigraphs; the ones that do are called circle graphs\footnote{From this definition it is not clear why these graphs are called \emph{circle} graphs; see for example~\cite{gioan2014practical} for more background.}.

\begin{figure}[t]
\begin{center}
\begin{tikzpicture}[scale=1.1] 
    \coordinate (v1) at (-2, 0);
    \coordinate (v2) at (-1, 0);
    \coordinate (v3) at (0, 0);
    \coordinate (v4) at (1, 0);
    \coordinate (v5) at (2, 0);

\draw[thick, green] (-2.2, 0) to[out=0, in=-180] (v1);

\draw[thick, black] (v5) to[out=0, in=-180] (2.2, 0);

\draw[thick, black] (v1) to[out=0, in=-180] (v2);

\draw[thick, green] (v2) to[out=0, in=-180] (v3);

\draw[thick, black] (v3) to[out=0, in=-180] (v4);

\draw[thick, green] (v4) to[out=0, in=-180] (v5);

\draw[thick, blue] (-2, -0.4) to[out=0, in=180+45] (v2);

\filldraw[white] (-1.5, -0.33) circle (2.5pt) {};

\draw[thick, red] (v1) to[out=-45, in=180+45] (v3);

\filldraw[white] (-0.5, -0.33) circle (2.5pt) {};

\draw[thick, blue] (v2) to[out=-45, in=180+45] (v4);

\filldraw[white] (0.5, -0.33) circle (2.5pt) {};

\draw[thick, red] (v3) to[out=-45, in=180+45] (v5);

\filldraw[white] (1.5, -0.33) circle (2.5pt) {};

\draw[thick, blue] (v4) to[out=-45, in=180] (2, -0.4);

\draw[thick, red] (v5) to[out=-45, in=180-20] (2.2, -0.15);

\draw[thick, red] (v1) to[out=180+45, in=20] (-2.2, -0.15);




    \filldraw[black] (v1) circle (2.5pt) node[above, font=\Large, yshift=5pt] {$0$};
    \filldraw[black] (v2) circle (2.5pt) node[above, font=\Large, yshift=5pt] {$1$};
    \filldraw[black] (v3) circle (2.5pt) node[above, font=\Large, yshift=5pt] {$2$};
    \filldraw[black] (v4) circle (2.5pt) node[above, font=\Large, yshift=5pt] {$3$};
    \filldraw[black] (v5) circle (2.5pt) node[above, font=\Large, yshift=5pt] {$4$};
\end{tikzpicture}
\vspace*{1mm}
\rule[0mm]{0.33\textwidth}{0.4pt} 
\vspace*{1mm}
\begin{tikzpicture}[scale=0.028] 
    \def\n{12}

    \def\radius{3cm}
    \def\inset{1.2cm} 

    \foreach \i in {0, 1, 2, 3, 4, 5, 6, 7, 8, 9, 10, 11} {
        \pgfmathsetmacro{\curradius}{\radius}
        \ifnum\i=0 \pgfmathsetmacro{\curradius}{\inset} \fi
        \ifnum\i=2 \pgfmathsetmacro{\curradius}{\inset} \fi
        \ifnum\i=4 \pgfmathsetmacro{\curradius}{\inset} \fi
        \ifnum\i=6 \pgfmathsetmacro{\curradius}{\inset} \fi
        \ifnum\i=8 \pgfmathsetmacro{\curradius}{\inset} \fi
        \ifnum\i=10 \pgfmathsetmacro{\curradius}{\inset} \fi
        \coordinate (v\i) at ({90 - 360 / \n * \i}:\curradius);
        \filldraw (v\i) circle (26pt) node[anchor={90 - 360 / \n * \i + 90}] {$\i$};
    }

    \foreach \i/\j in {
        0/2, 2/4, 4/6, 6/8, 8/10, 10/0} {
        \draw[red, thick] (v\i) -- (v\j);
    }
        \foreach \i/\j in {
        2/1, 4/3, 6/5, 8/7, 10/9, 11/0} {
        \draw[green, thick] (v\i) -- (v\j);
    }

        \foreach \i/\j in {
        1/3, 3/5, 5/7, 7/9, 9/11, 11/1} {
        \draw[blue, thick] (v\i) -- (v\j);
    }

    \foreach \i/\j in {
        3/2, 5/4, 7/6, 9/8, 11/10,
        11/10, 1/0} {
        \draw[thick] (v\i) -- (v\j);
    }
\end{tikzpicture}
\end{center}
\vspace*{-3mm}
\caption{\justifying Top: A segment of a $4$-regular graph associated with a cycle graph. The general case repeats periodically. Bottom: An explicit example of the $4$-regular multigraph associated with the $12$-cycle. An Eulerian tour corresponding to the $12$-cycle is shown by following the edges in the ordering of red, green, blue, and black/$0, 2, 1, 3, 2$, etc. The inner vertices have been moved to the center to highlight the fact that these will be measured out. }
\label{fig:circulant_graph}
\end{figure}
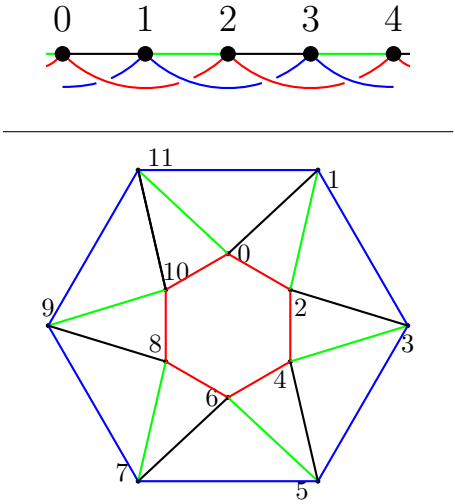

The $4$-regular multigraph associated with a cycle graph is particularly simple---it is given by the \emph{circulant graph} $C_n^{1, 2}$, i.e., it is the multigraph with vertices $1$ to $n$, where $i, j$ are connected iff $\left|i-j\right|\leq 2 \textrm{ mod }n$. To see this, we depict the edges of the associated Eulerian tour associated with a $12$-cycle in Fig.~\ref{fig:circulant_graph}, indicated by following the edges in the ordering of red, green, blue and black. The general case follows straightforwardly.

\begin{figure}[t]
\begin{center}
\begin{tikzpicture}[scale=0.06] 
    \def\n{12}

    \def\radius{1.5cm}
    \def\inset{0.6cm} 

    \foreach \i in {0, 1, 2, 10, 11} {
        \pgfmathsetmacro{\curradius}{\radius}
        \ifnum\i=0 \pgfmathsetmacro{\curradius}{\inset} \fi
        \ifnum\i=2 \pgfmathsetmacro{\curradius}{\inset} \fi
        \ifnum\i=10 \pgfmathsetmacro{\curradius}{\inset} \fi
        \coordinate (v\i) at ({90 - 360 / \n * \i}:\curradius);
    }

    \foreach \i/\j in {
        0/2, 0/10, 1/11, 0/11,  0/1} {
        \draw[thick] (v\i) -- (v\j);
    }

    \foreach \i in {1, 11}{
        \filldraw (v\i) circle (23pt);
}


\filldraw[fill=black, draw=black] (v0) circle (37pt);
\end{tikzpicture}
\vspace*{3mm}
\rule[0mm]{0.33\textwidth}{0.4pt} 
\vspace*{1mm}
\end{center}
\begin{subfigure}[b]{0.5\textwidth}
\begin{center}
\begin{tikzpicture}[scale=1.0] 
        \coordinate (vlt) at (-1, 1);
        \coordinate (vrt) at (1, 1);

\coordinate (vlb) at (-1, 0);
        \coordinate (vrb) at (1, 0);

\draw[thick, red, dotted, line width=0.5mm] (vlt) to[out=45, in=180-45] (vrt);
\draw[thick, red, dotted, line width=0.5mm] (vlt) to[out=-45, in=-180+45] (vrt);

\draw[thick, red, dotted, line width=0.5mm] (vlb)--(vrb);

        \node[scale=0.41, draw, circle, line width=0.5mm, fill=black] (7) at (vlt) {};

        \node[scale=0.41, draw, circle, fill=black] (7) at (vrt) {};



\end{tikzpicture}
\begin{tikzpicture}[scale=1.0] 
          \coordinate (vlt) at (-1, 1);
        \coordinate (vrt) at (1, 1);

\coordinate (vlb) at (-1, 0);
        \coordinate (vrb) at (1, 0);

\draw[thick, blue] (vrt) to (vlb);
\node[scale=0.81, circle, fill=white] (9) at (0, 0.5) {};
\draw[thick, blue] (vlt) to (vrb);

\draw[thick, blue] (vlt) to[out=0, in=-180] (vrt);


        \node[scale=0.41, draw, circle, fill=black] (7) at (vlt) {};

        \node[scale=0.41, draw, circle, fill=black] (7) at (vrt) {};



\end{tikzpicture}
\begin{tikzpicture}[scale=1.0] 
          \coordinate (vlt) at (-1, 1);
        \coordinate (vrt) at (1, 1);

\coordinate (vlb) at (-1, 0);
        \coordinate (vrb) at (1, 0);

\draw[thick, blue!15!green!76, dashed, line width=0.5mm] (vlt) to[out=-30, in=0] (vlb);
\draw[thick, line width=0.5mm, blue!15!green!76, dashed] (vrt) to[out=180+30, in=180] (vrb);

\draw[thick, , line width=0.5mm, blue!15!green!76, dashed] (vlt) to[out=0, in=-180] (vrt);

        \node[scale=0.41, draw, circle, fill=black] (7) at (vlt) {};

        \node[scale=0.41, draw, circle, fill=black] (7) at (vrt) {};



\end{tikzpicture}
\end{center}
\end{subfigure}
\vspace*{1mm}
\caption{\justifying Top: a small fragment of the graph in Fig.~\ref{fig:circulant_graph}. Bottom: the three possible transition minors of the above fragment. We will refer to these as $X$-, $Y$-, and $Z$-fragments, respectively. The colors used here are unrelated to the colors used in Fig.~\ref{fig:circulant_graph}.}
\label{fig:transition_minor}
\end{figure}
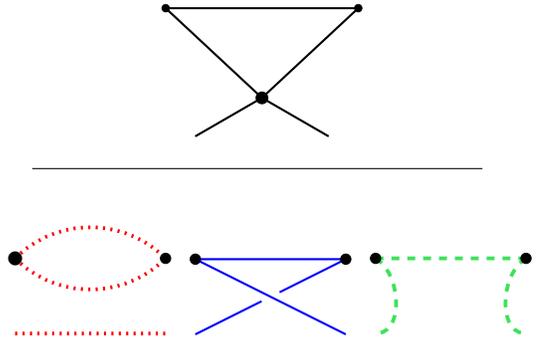

Pauli measurements on (elements of an equivalence class) on circle graphs have a simple interpretation in terms of $4$-regular multigraphs. The three Pauli measurements on a vertex $v$ of a $4$-regular graph yield three possible \emph{transition-minors}, which are $4$-regular graphs on the same vertex set as $F$ but with $v$ removed. Furthermore, the three different transition-minors correspond to the three different ways to `disconnect' the four edges incident on $v$, see Fig.~\ref{fig:transition_minor}. Transition-minors will be useful to us since the Eulerian tours on the three transition-minors on $v$ give exactly the possible vertex-minors when measuring out $v$. For more information on transition-minors and vertex-minors of circle graphs, see section F of~\cite{dahlberg2020counting}, or definition C.1.3 from~\cite{dahlberg2021entanglement}.

In the top of Fig.~\ref{fig:transition_minor}, we consider a small fragment of $C_n^{1,2}$. Note that this fragment repeats. The bottom of Fig.~\ref{fig:transition_minor} shows the three possible transition minors of that fragment. We label these as the $X$-, $Y$-, and $Z$-fragments, since they correspond to $X$-, $Y$-, and $Z$-measurements with respect to the Eulerian tour from Fig.~\ref{fig:circulant_graph}.

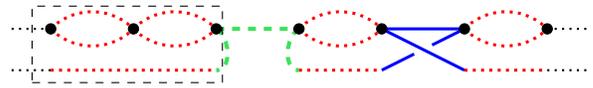
\begin{figure}[t]
\begin{center}
 \begin{tikzpicture}[scale=0.55] 

  \foreach \i in {1, 2, 3, 4, 5, 6, 7} {
        \coordinate (vt\i) at (-1+2*\i, 1);
                
\coordinate (vb\i) at (-1+2*\i, 0);

}

\draw[line width=0.4mm, red, dotted] (vt1) to[out=45, in=180-45] (vt2);
\draw[line width=0.4mm, red, dotted] (vt1) to[out=-45, in=-180+45] (vt2);

\draw[line width=0.4mm, red, dotted] (vb1)--(vb2);


\draw[line width=0.4mm, red, dotted] (vt2) to[out=45, in=180-45] (vt3);
\draw[line width=0.4mm, red, dotted] (vt2) to[out=-45, in=-180+45] (vt3);

\draw[line width=0.4mm, red, dotted] (vb2)--(vb3);


\draw[line width=0.55mm, blue!15!green!76, dashed] (vt3) to[out=-30, in=0] (vb3);
\draw[line width=0.55mm, blue!15!green!76, dashed] (vt4) to[out=180+30, in=180] (vb4);

\draw[line width=0.55mm, blue!15!green!76, dashed] (vt3) to[out=0, in=-180] (vt4);


\draw[line width=0.4mm, red, dotted] (vt4) to[out=45, in=180-45] (vt5);
\draw[line width=0.4mm, red, dotted] (vt4) to[out=-45, in=-180+45] (vt5);

\draw[line width=0.4mm, red, dotted] (vb4)--(vb5);


\draw[line width=0.4mm, blue] (vt6) to (vb5);
\node[scale=0.81, circle, fill=white] (9) at (-1+2*6-1, 0.5) {};
\draw[line width=0.4mm, blue] (vt5) to (vb6);

\draw[line width=0.4mm, blue] (vt5) to[out=0, in=-180] (vt6);


\draw[line width=0.4mm, red, dotted] (vt6) to[out=45, in=180-45] (vt7);
\draw[line width=0.4mm, red, dotted] (vt6) to[out=-45, in=-180+45] (vt7);

\draw[line width=0.4mm, red, dotted] (vb6)--(vb7);


  \foreach \i in {1, 2, 3, 4, 5, 6, 7} {
        \node[scale=0.41, draw, circle, fill=black] (7) at (vt\i) {};
            }


\draw[thick, dotted] (vt1) to (-0, 1);
\draw[thick, dotted] (vb1) to (-0, 0);

\draw[thick, dotted] (vt7) to (14, 1);
\draw[thick, dotted] (vb7) to (14, 0);

  \draw[dashed] (0.55, -0.3) -- (5.15, -0.3) -- (5.15, 1.55) -- (0.55, 1.55) -- cycle;

\end{tikzpicture}   
\end{center}
\vspace{-1mm}
\caption{\justifying An example of a transition-minor of a circulant graph (when measuring out only even qubits). In general, a transition-minor will consist of a similar sequence of $X$-, $Y$-, and $Z$-fragments (see Fig.~\ref{fig:transition_minor}).}
\label{fig:transition_minor_total}
\end{figure}

From the above, we find that the possible transition-minors are all possible sequences of length $n$ of $X$-, $Y$-, and $Z$-fragments (starting from a fixed vertex), see Fig.~\ref{fig:transition_minor_total}. We will write such a sequence $S$ as a word of length $n$ from the alphabet $\lbrace{X, Y, Z\rbrace}$.

Let us briefly recapitulate. We moved to the $4$-regular multigraph picture since understanding all possible post-measurement states (up to single-qubit Clifford rotations) corresponds to understanding all possible Eulerian tours on certain transition minors of $C_n^{1,2}$. We have shown that the transition minors can be described by a (cyclic) word $S$. Since we do not care about the post-measurement state up to local single-qubit Cliffords, we will find instead only one Eulerian tour for each word $S$.

First, we can restrict ourselves to subwords that do not contain $Z$. This follows from the fact that a $4$-regular multigraph of two disjoint circle graphs corresponds to a \emph{connected sum} of the $4$-regular multigraphs of the individual circle graphs, see section 1 of~\cite{traldi2015transition}. Intuitively, this follows from the fact that a $Z$ measurement corresponds to vertex deletion on the cycle graph. 

\begin{figure}[t]
\vspace*{4mm}
\begin{center}
    \begin{tikzpicture}[scale=0.7] 
    \coordinate (v1) at (0, 0); 

\draw[thick, red] (v1) to (-1, -1);
\draw[thick, green] (v1) to (-1, 1);
\draw[thick, blue] (v1) to (1, 1);
\draw[thick, black] (v1) to (1, -1);

    \filldraw[black] (v1) circle (2.5pt) node[above, font=\Large, yshift=5pt] {$v$};

    \coordinate (v2) at (3, 0); 
        \coordinate (v3) at (4, 0); 

\draw[thick, red] (v2) to (-1+3, -1);
\draw[thick, green] (v2) to (-1+3, 1);
\draw[thick, blue] (v3) to (1+4, 1);
\draw[thick, black] (v3) to (1+4, -1);

\draw[thick, magenta] (v2) to[out=-30, in=180+30] (v3);
\draw[thick, magenta] (v2) to[out=30, in=180-30] (v3);

    \filldraw[black] (v1) circle (2.5pt) node[above, font=\Large, yshift=5pt] {$v$};

\node[rectangle, text=black, font=\Large] at (1.55, 0) {$\mapsto$};

    \filldraw[black] (v2) circle (2.5pt) node[above, font=\Large, yshift=5pt] {$v_1$};

        \filldraw[black] (v3) circle (2.5pt) node[above, font=\Large, yshift=5pt] {$v_2$};

    \coordinate (vt2) at (0.95, -3); 
        \coordinate (vt3) at (1.95, -3); 
                \coordinate (vt4) at (2.95, -3); 

\draw[thick, red] (vt2) to ([shift=({-1cm,-1cm})]vt2);
\draw[thick, green] (vt2) to ([shift=({-1cm,1cm})]vt2);
\draw[thick, blue] (vt4) to ([shift=({1cm,1cm})]vt4);
\draw[thick, black] (vt4) to ([shift=({1cm,-1cm})]vt4);

\draw[thick, magenta] (vt2) to[out=-30, in=180+30] (vt3);
\draw[thick, magenta] (vt2) to[out=30, in=180-30] (vt3);
\draw[thick, magenta] (vt3) to[out=-30, in=180+30] (vt4);
\draw[thick, magenta] (vt3) to[out=30, in=180-30] (vt4);


\node[rectangle, text=black, font=\Large] at (-1, -3) {$\mapsto$};

    \filldraw[black] (vt2) circle (2.5pt) node[above, font=\Large, yshift=5pt] {$v_1$};
    \filldraw[black] (vt3) circle (2.5pt) node[above, font=\Large, yshift=5pt] {$v_2$};
    \filldraw[black] (vt4) circle (2.5pt) node[above, font=\Large, yshift=5pt] {$v_3$};

\end{tikzpicture}
\end{center}
\caption{\justifying An example of a leaf expansion on a $4$-regular multigraph with respect to $v$, and an Eulerian tour corresponding to the edges red, green, blue, and black. For simplicity, we do not show any other part of the graph. Note that the graph on the right inherits an Eulerian tour from $U$, given by the edges red, the two edges between $v_1$ and $v_2$ in succession, green, blue, and black.}
\label{fig:leaf_expansion}
\end{figure}
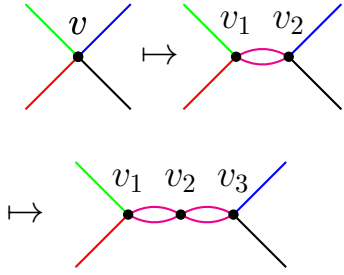

For the second part, we introduce \emph{leaf expansions}. A leaf expansion maps a given $4$-regular graph $F$ and constructs a $4$-regular multigraph $F'$ with one more vertex. To define a leaf expansion on $F$, we will need an Eulerian tour $U$ on $F$ and a vertex $v$. The leaf expansion of $F$ with respect to $U$ and $v$ is a $4$-regular multigraph $F'$ with vertex set $V(F) \Delta \lbrace{v, v_1, v_2\rbrace}$, where $\Delta$ is the symmetric set difference operation. That is, we can think of `splitting' the vertex $v$ into two vertices $v_1, v_2$. Furthermore, the edge set of $F'$ is the same as $F$ (respecting the relabelling of $v$ to $v_1$, $v_2$), and with two additional multi-edges between $v_1$ and $v_2$. We show a leaf expansion in Fig.~\ref{fig:leaf_expansion}; note that $F'$ inherits an Eulerian tour $U'$ from $U$, as indicated. Note that technically speaking a leaf expansion is not well-defined; there is an arbitrary choice of which vertex will be $v_1$ and which one $v_2$. However, both resultant graphs are isomorphic. 

The terminology of a leaf expansion comes from the following fact: if $G$ is the graph associated with the Eulerian tour, then the corresponding leaf expansion $F'$ has an Eulerian tour $U'$ (see Fig.~\ref{fig:leaf_expansion}) on it such that the associated graph is isomorphic to $G$, but with a leaf attached to the vertex corresponding to $v$. The proof of this fact follows from directly writing out the sequence of vertices visited from $U'$.

\begin{figure}[t]
\begin{center}
    \begin{tikzpicture}[scale=0.85] 
          \coordinate (vt1) at (-1, 1);
        \coordinate (vt2) at (1, 1);                
\coordinate (vb1) at (-1, 0);
        \coordinate (vb2) at (1, 0);
        
\coordinate (vt3) at (3, 1);
        \coordinate (vb3) at (3, 0);

\coordinate (vt4) at (5, 1);
        \coordinate (vb4) at (5, 0);

\draw[thick, blue] (vt2) to (vb1);
\node[scale=0.81, circle, fill=white] (9) at (0, 0.5) {};
\draw[thick, blue] (vt1) to (vb2);
\draw[thick, blue] (vt1) to[out=0, in=-180] (vt2);

\draw[thick, blue] (vt3) to (vb2);
\node[scale=0.81, circle, fill=white] (9) at (2, 0.5) {};
\draw[thick, blue] (vt2) to (vb3);
\draw[thick, blue] (vt2) to[out=0, in=-180] (vt3);

\draw[thick, blue] (vt4) to (vb3);
\node[scale=0.81, circle, fill=white] (9) at (4, 0.5) {};
\draw[thick, blue] (vt3) to (vb4);
\draw[thick, blue] (vt3) to[out=0, in=-180] (vt4);

        \node[scale=0.41, draw, circle, fill=black] (7) at (vt1) {};

        \node[scale=0.71, draw, circle, fill=black] (7) at (vt2) {};

        \node[scale=0.71, draw, circle, fill=black] (7) at (vt3) {};

        \node[scale=0.41, draw, circle, fill=black] (7) at (vt4) {};

\node[rectangle, text=black, font=\Large, rotate=-90] at (2, -0.5) {$\mapsto$};

\end{tikzpicture}
    \begin{tikzpicture}[scale=1.0] 
          \coordinate (vt1) at (-1, 1);
        \coordinate (vt2l) at (0.5, 1);
        \coordinate (vt2r) at (1.5, 1);          
\coordinate (vb1) at (-1, 0);
        \coordinate (vb2) at (1, 0);
        
\coordinate (vt31) at (3-0.66, 1);
\coordinate (vt32) at (3, 1);
\coordinate (vt33) at (3+0.66, 1);

        \coordinate (vb3) at (3, 0);

\coordinate (vt4) at (5, 1);
        \coordinate (vb4) at (5, 0);

\draw[thick, blue] (vt2l) to (vb1);
\node[scale=0.81, circle, fill=white] (9) at (-0.15, 0.565) {};
\draw[thick, blue] (vt1) to (vb2);
\draw[thick, blue] (vt1) to[out=0, in=-180] (vt2l);

\draw[thick, red, dotted, line width=0.5mm] (vt2l) to[out=45, in=180-45] (vt2r);
\draw[thick, red, dotted, line width=0.5mm] (vt2l) to[out=-45, in=-180+45] (vt2r);

\draw[thick, blue] (vt31) to (vb2);
\node[scale=0.81, circle, fill=white] (9) at (1.95, 0.685) {};
\draw[thick, blue] (vt2r) to (vb3);
\draw[thick, blue] (vt2r) to[out=0, in=-180] (vt31);

\draw[thick, red, dotted, line width=0.5mm] (vt31) to[out=45, in=180-45] (vt32);
\draw[thick, red, dotted, line width=0.5mm] (vt31) to[out=-45, in=-180+45] (vt32);
\draw[thick, red, dotted, line width=0.5mm] (vt32) to[out=45, in=180-45] (vt33);
\draw[thick, red, dotted, line width=0.5mm] (vt32) to[out=-45, in=-180+45] (vt33);

\draw[thick, blue] (vt4) to (vb3);
\node[scale=0.81, circle, fill=white] (9) at (4.198, 0.587) {};
\draw[thick, blue] (vt33) to (vb4);
\draw[thick, blue] (vt33) to[out=0, in=-180] (vt4);

        \node[scale=0.41, draw, circle, fill=black] (7) at (vt1) {};

        \node[scale=0.71, draw, circle, fill=black] (7) at (vt2l) {};
        \node[scale=0.71, draw, circle, fill=black] (7) at (vt2r) {};

                \node[scale=0.71, draw, circle, fill=black] (7) at (vt31) {};
        \node[scale=0.71, draw, circle, fill=black] (7) at (vt32) {};
        \node[scale=0.71, draw, circle, fill=black] (7) at (vt33) {};
        
        \node[scale=0.41, draw, circle, fill=black] (7) at (vt4) {};

\end{tikzpicture}
\end{center}
\caption{\justifying Example of how a sequence of leaf expansions on a $4$-regular graph of a cycle graph/path graph corresponds to a sequence of $X$- and $Y$-fragments.}
\label{fig:leaf_expansion_on_fragment}
\end{figure}
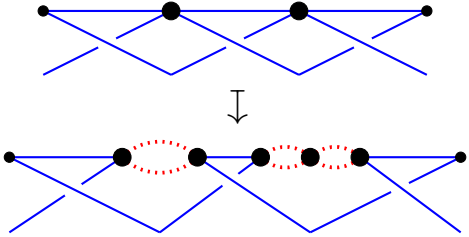

From Fig.~\ref{fig:leaf_expansion_on_fragment}, it is clear that any segment consisting of $X$- and $Y$-fragments corresponds to a sequence of leaf expansions on a path graph. In other words, the possible states are (locally equivalent to) disjoint copies of caterpillar graph states, where the vertices that are part of the subgraphs corresponding to star graphs need to be contingent. Furthermore, there is one final possibility where no $Z$-measurement is performed. In that case, one ends up with a cycle graph with attached leaves (with the same restriction on contingency).

Finally, let us say something about the path graph shown in Fig.~\ref{fig:FusingBuildingBlocks}a, where the server owns every third qubit, i.e., every third qubit will be measured. Since this graph is a combination of zigzag elements, each of the measured qubits has an identical neighborhood to the zigzag graph. Allowing the same measurements as before, starting from Fig.~\ref{fig:FusingBuildingBlocks}a, we can only distribute disjoint copies of caterpillar graph states where each vertex of the path graph can have one leaf maximum.

\bibliography{IEEEabrv,Literature}{}

\end{document}